\documentclass[12pt]{iopart}
\usepackage{graphicx}
\usepackage{epsfig}
\usepackage{epstopdf}
\usepackage{verbatim}

\begin{document}
\title[Thwarting the PNS Attack with Entanglement Enhanced BB84]{Thwarting the Photon Number Splitting Attack with Entanglement Enhanced BB84 Quantum Key Distribution}
\author{Carl F Sabottke$^1$, Chris D Richardson$^1$, Petr M Anisimov$^1$, Ulvi Yurtsever$^{1,2}$, Antia Lamas-Linares, Jonathan P Dowling$^1$}
\address{$^1$ Hearne Institute for Theoretical Physics, Department of Physics and Astronomy, Louisiana State University, Baton Rouge, LA 70803}
\address{$^2$ MathSense Analytics 1273 Sunny Oaks Circle Altadena, CA 91001}
\ead{csabot3@lsu.edu}

\begin{abstract}
We develop an improvement to the weak laser pulse BB84 scheme for quantum key distribution, which utilizes entanglement to improve the security of the scheme and enhance its resilience to the photon-number-splitting attack.  This protocol relies on the non-commutation of phase and number to detect an eavesdropper performing quantum non-demolition measurement on photon number.  The potential advantages and disadvantages of this scheme are compared to the coherent decoy state protocol.
\end{abstract}


\maketitle{}

\section{Introduction}

Quantum key distribution is rapidly emerging as an elegant application of quantum information theory with immense practical value.  The advent of quantum computing compromises classical encryption schemes which are dependent on computational difficulty for security.  Fortunately, quantum information theory solves the exact problem it creates.  If a transmitter, Alice, wants to exchange a message with a receiver, Bob, then the fundamental principles of quantum mechanics allow them to generate a key that cannot be obtained by an eavesdropper, Eve [1\--3].

In the theoretic framework of BB84, Alice sends a sequence of single photon pulses to Bob.  These photons are prepared in randomly chosen orthogonal bases.  In the receiving lab, Bob has two bases in which to measure the photon and randomly alternates between them.  If Eve tries measuring Alice's photon and then sending the result of her measurement to Bob, the eavesdropper will introduce errors into the key, since she does not know in which basis the photon is being sent nor does she know in which basis Bob will measure.  Alice and Bob can then use these errors to detect the eavesdropper's presence and determine the security of the key [4].

However, in many experimental settings, Alice does not have a true single photon source, so she sends weak laser pulses (WLP) instead.  This coherent light photon number probability follows a Poisson distribution.  The probability of a pulse containing $n$ photons is \begin{equation}
P_{n} = \frac{\mu ^n}{n! e^\mu}
\end{equation}
where $\mu$ is the mean photon number which will be taken to be a positive number less than one to avoid pulses with more than one photon.  However, multiple photon pulses will still occur with probability $P_{M} = 1-e^{-\mu} - {\mu}e^{-\mu}$.  This exposes the scheme to the photon number splitting (PNS) attack.

To perform the PNS attack, Eve replaces the high loss channel that Alice and Bob are using with a lossless channel.  Eve then performs a quantum non-demolition (QND) measurement on each pulse to obtain number information without perturbing the bases in which the information is encoded.  When she determines a pulse with a single photon is in the line, Eve simulates the loss of the original line by blocking a fraction of these pulses.  When Eve observes a pulse that has multiple photons, she splits the pulse and stores a photon in a quantum memory.  Eve then sends the rest of the pulse to Bob.  After Alice and Bob perform public discussion and announce the bases used for each pulse, Eve can retrieve the photons from her quantum memory and obtain a significant fraction of the key without being detected by Alice and Bob [5\--9].

In general, all losses must be attributed to eavesdropping and privacy amplification methods are used to distill a smaller secret key from the raw key generated via the BB84 protocol. In single photon BB84, the distilled secure key rate has approximately linear dependence on the transmittivity.  However, for WLP BB84, the PNS attack reduces the secure key rate to approximately quadratic dependence on the channel's transmittivity [10].  In a typical high loss situation, this presents a major problem for the key rate. One solution is to use coherent decoy states, a technique which has met with multiple experimental successes [11\--18].  Another alternative is to use entanglement to effectively trump Eve's use of the PNS attack.  This is the impetus for the development of our entanglement enhanced scheme for BB84.  For convenience and clarity, we will refer to this entanglement enhanced WLP BB84 as EE BB84.
 
Most entanglement based quantum key distribution schemes rely on violations of Bell's inequalities to ensure security [19].  However, this is not the strategy that our EE BB84 employs here.  Instead, we detect Eve by introducing an entangled quantum state into the system that is not used to transmit key bits but only to detect Eve's QND measurements. In figure 1 we schematically illustrate how such an entanglement ancilla may be generated. This allows for a recovery of an approximately linear dependence on transmittivity for the key rate.  EE BB84 shares this advantage with coherent decoy state protocols as well as schemes that utilize strong phase reference pulses to eliminate Eve's ability to send Bob vacuum signals [10].

\section{EE BB84 Scheme}

\begin{figure}
	\centering
	\includegraphics[width=1\linewidth]{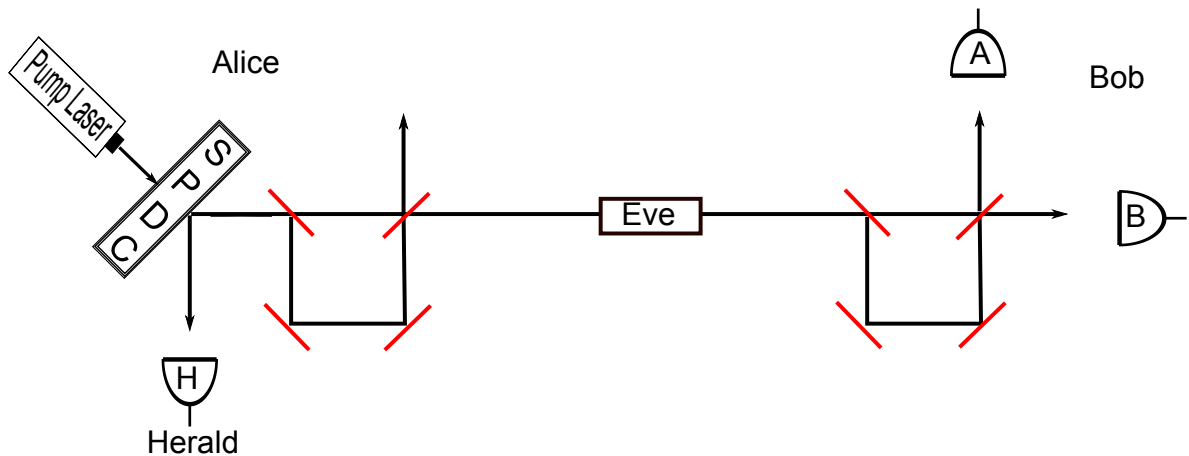}
	\caption{In the entanglement ancilla, for each photon pair generated by Alice, one is detected in her lab to obtain time information.  The other is sent into a beamsplitter and then recombined at a second beamsplitter in the lab to create a pulse with halves that have a time delay that results from a length difference in the paths between the two beamsplitters.  This pulse is then sent through the channel to Bob, who passes the pulse through two beamsplitters in his lab that have path differences identical to those in Alice's lab.}
	\label{fig:Chernoff_experiment}
\end{figure}

In our EE BB84, Alice and Bob randomly alternate between implementing WLP BB84 and an entangled decoy state ancilla.  The entangled states are not primarily used to distribute key bits.  Instead, Alice and Bob use the entangled states to detect the presence of an eavesdropper.  Alice sends the entangled pulses randomly mixed with the weak laser pulses to guard against the use of a QND measurement device.  When Eve measures photon number in the PNS attack on unaugmented WLP BB84, she avoids detection. The QND measurement collapses the coherent state into a number state, which Bob cannot distinguish from the coherent state.  This is related to the fact that the number operator commutes with the prepared bases.  However, phase and number do not commute, as they are conjugate variables.  Therefore, Alice and Bob can use the phase information provided by phase entangled decoy states to detect Eve whenever she chooses an attack scheme that involves measuring number.

In the entangled state mode, we generate two time-entangled photons using spontaneous parametric down conversion (SPDC).  Alice measures one photon in the pair to obtain an accurate time of emission for the other photon.  This combination of pump laser, SPDC, and detection of one of the pair of photons gives us a heralded single photon source. As in BB84, the heralded photon is randomly assigned either a horizontal, vertical, diagonal, or anti-diagonal polarization.  Then, the heralded photon is sent to a beam splitter which leads to the state $| \Psi \rangle = \frac{1}{\sqrt{2}}(|10 \rangle + |01 \rangle )$.  Half of the state travels down the longer arm, while the other half travels down the shorter arm.  The halves recombine at the second beam splitter where there is a probability for the state to leave the quantum channel (see figure 1).  A detector will distinguish these possibilities and allows them to be ignored.  However, when the pulse does exit into the quantum channel, it is an entangled pulse, where half is delayed in time due to extra path length of the long arm.

\begin{figure}
	\centering
	\includegraphics[width=.7\linewidth]{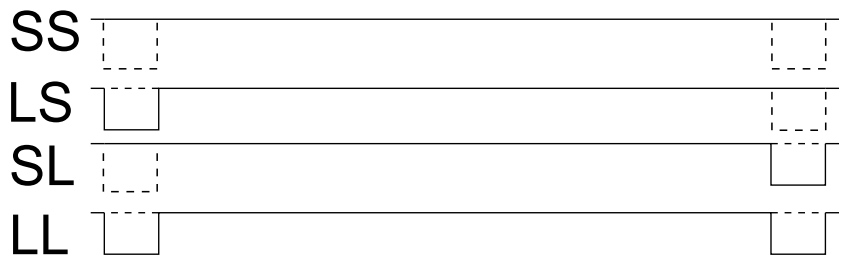}
	\caption{Possible paths a photon can take to get from Alice to Bob's detector:  short-short (SS), long-long (LL), short-long (SL), and long-short (LS).  Alice's time information allows the SL and LS paths, which are indistinguishable from each other to be distinguished from both SS and LL.}
	\label{fig:Chernoff_experimentpaths}
\end{figure}

When Bob receives the test pulse from Alice in his lab, he detects the pulse by sending it through a beam splitter which puts the pulse through long and short arms identical to the setup in Alice's lab.  The pulse then encounters the final beam splitter.  In this process, there are three possibilities for the pulse.  The strong time information from the photon initially detected by Alice allows for the differentiation between these three outcomes.  One possibility is that the photon takes the short path both times, labeled SS in figure \ref{fig:Chernoff_experimentpaths}.  Another outcome is that the photon takes the long path both times, labeled as LL.  These two possibilities do not yield strong information about Eve's activities.  However, the other possibility is that the photon travels down one long path and one short path, labeled LS or SL.  This possibility can detect the use of a quantum non-demolition measurement device \cite{bib:DowlingQND}. The photon's self-interference will result in a bright port and a dark port in Bob's detection apparatus.  Yet, if Eve is measuring number for the PNS attack, then Bob's dark port will not be completely dark.  Obviously, it will not be completely dark even without an eavesdropper, since a practical system will have imperfections and not identically match the ideal case. Nevertheless, Eve's actions will still introduce additional error, which can be used to detect her presence.

In our setup, Bob's detection scheme for the entangled pulses is different from his detection scheme for the signal states.  This is less than ideal, because if the mode that Alice and Bob are operating in at any given time is not random, then the security of the entire protocol is compromised.  If Eve can predict whether a signal state or a decoy state is being sent, then she can adjust her attack plan accordingly and render the entangled states useless.  Therefore, it is critical that Eve cannot distinguish between the entangled states and the signal states.  Additionally, Alice and Bob must randomly alternate between the signal and decoy modes.  Felicitously, the decoy mode does not need to be run with very high frequency in order to detect the use of a quantum non-demolition attack.  Nevertheless, since Alice and Bob must each run separate modes for the signal states and the decoy states, a fraction of the pulses they exchange will be worthless.  Alice and Bob runs WLP BB84 protocol with frequencies $f_{SA}$ and $f_{SB}$ respectively.  They implement the entangled state decoy ancilla with frequencies $f_{DA}$ and $f_{DB}$.  Alice and Bob exchange key information with frequency $f_{SA}f_{SB}+f_{DA}f_{SB}$, and the entangled decoy pulses yield information about the presence of a quantum non-demolition measurement device with frequency $f_{DA}f_{DB}$.  With frequency $f_{SA}f_{DB}$, Alice and Bob are operating in incompatible modes, and these exchanges will provide no valuable information, because Bob does not obtain polarization information when measuring phase.  Since $f_{SA}$ and $f_{SB}$ are much larger than $f_{DA}$ and $f_{DB}$, this inefficiency is undesirable, but ultimately does not significantly diminish the practicality of the scheme.  Nevertheless, it is also indicative of the trade-off in quantum cryptography between speed and security.

\section{Symmetric Hypothesis testing and the Chernoff Distance}

We use Chernoff distance \cite{bib:Chernoff} and symmetric hypothesis testing to calculate the confidence in which Eve is known to be listening or not listening \cite{bib:tripwire}.  For EE BB84 the null hypothesis is that Eve is not measuring number using a QND measurement device, and the alternative hypothesis is that Eve is using such a device to measure number.  For the null hypothesis, the probability that the photon will enter the bright port is $p$, and there is $\overline{p} = 1 - p$ probability for the photon to enter the dark port.  When Eve is acting on the system in the alternative hypothesis, there is a probability $q$ for the photon to enter Bob's light port and a probability $\overline{p} = 1 - p$ for it to enter the dark port.  Furthermore, the maximum probability $P_{\mathrm{Error}}^{\mathrm{Max}}$ of a false positive or of choosing the wrong hypothesis after $n$ trials is:

\begin{equation}
	P_{\mathrm{Error}}^{\mathrm{Max}} = \frac{1}{2}e^{-n C(p,q)}
	\label{eqn:Error_probability}
\end{equation}
where $C(p,q)$ is the Chernoff distance given by the equation:
\begin{equation}
	C(p,q) = \xi \mathrm{ln}(\frac{\xi}{p}) + \overline{\xi} \mathrm{ln}(\frac{\overline{\xi}}{\overline{p}})
	\label{eqn:ChernoffDistance}
\end{equation}
where $\xi = \frac{\mathrm{ln}(\frac{\overline{q}}{\overline{p}})}{\mathrm{ln}(\frac{p}{\overline{p}})+\mathrm{ln}(\frac{\overline{q}}{q})}$ and $\overline{\xi} = 1 - \xi$.

We use equations \ref{eqn:Error_probability} and \ref{eqn:ChernoffDistance} to calculate the number of trials needed for a given maximum uncertainty $P_{Error}^{Max}$:
\begin{equation}
	n = \frac{-\mathrm{ln}(2 P_{\mathrm{Error}}^{\mathrm{Max}})}{C(p,q)}.
	\label{eqn:photon_number}
\end{equation}

This analysis determines the number of trials necessary for a given confidence of detecting an eavesdropper for EE BB84 and coherent decoy states.

\section{EE BB84 Statistical Analysis}

In an ideal scenario, with no dephasing from the environment, we can easily construct the probabilities of the two hypotheses.  For the null hypothesis, the probability that the photon will enter the bright port is $p=1$, and there is $\overline{p} = 1 - p = 0$ probability for the photon to enter the dark port.  When Eve is acting on the system in the alternative hypothesis, there is an equal probability, $q = \overline{q}  = \frac{1}{2}$, for the photon to enter either of Bob's detectors.  This results in a Chernoff distance of .69.  Therefore, if we define a trial to be a photon sent from Alice and detected by Bob, the number of trials to detect Eve at the 99\% confidence level ($P_{\mathrm{Error}}^{\mathrm{Max}} = 0.01$) requires an exchange of a maximum of just 6 photons between Alice and Bob.

We are only investigating the photons that reach Bob with the proper time information.  Thus, unlike the coherent decoy states, loss is not the most significant quantity to investigate quantitatively.  Instead, dephasing is our primary concern.  The environment can affect the entangled decoy state by changing the phase information in it.  Since the two states are sent down the line close together, it might be assumed that any environmental factor that would affect one half of the state, would affect the other and therefore the total phase information in the state would remain unchanged.  However, since in our framework, dephasing is what would affect the scheme the most, we still want to investigate its effect on the Chernoff distance.

\begin{figure}[ht]
	\centering
	\includegraphics[scale=1]{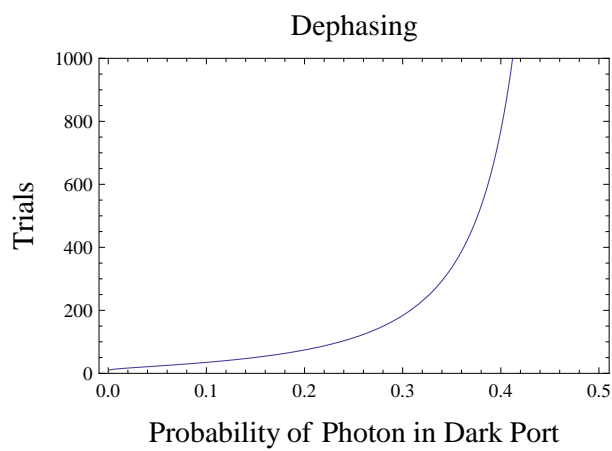}
	\caption{Dephasing can be caused by the environment, an eavesdropper or both.  As the dephasing increases, the probability of finding a photon in the dark port increases. This causes the number of trials needed to detect an eavesdropper with a 99\% confidence to increase.  When the probability of detecting a photon at the light and dark port is equal, it becomes impossible to tell an eavesdropper apart from the environment.}
	\label{fig:number_vs_loss}
\end{figure}

When dephasing is included, the problem turns into that of determining whether a coin is fair.  The question becomes: how many trials does it take to be confident that Eve is there or not? When dephasing is present, the probability for a photon to be detected in the dark port increases.  It becomes more difficult to tell Eve apart from the environment.  With complete dephasing the probability to find a photon in either the bright port or the dark port becomes 50-50.  Figure \ref{fig:number_vs_loss} shows how many trials are needed to have a 99\% confidence of determining if Eve is listening or not versus the probability of finding a photon in the dark port (dephasing) regardless of Eve.

\section{Coherent Decoy States Statistical Analysis}

The alternative to EE BB84 is the popular coherent decoy state solution.   In the PNS attack, Eve assumes Alice's photon source has a constant mean photon number.   However, if Alice randomly alters the mean photon number of her source in a way that is known to her, but not perceivable to Eve, then she can detect the PNS attack.  This is the idea that motivates coherent decoy states.  Pulses from the source with a higher mean photon number will contain a greater fraction of multi-photon pulses, which Eve will not block.  Therefore, when Alice and Bob discuss the protocol, Alice can compare the loss in the line for when different mean photon numbers were used.  If there is a marked difference between the loss for the decoy states and the loss for the signal states, then Alice can conclude that Eve is using the photon number splitting attack [22\--26].

We treat coherent decoy states in a similar manner to EE BB84, but instead of dephasing being the key quantity of interest, loss is, because Eve hides in the loss of the system.  The coherent decoy state solution uses two (or more) attenuated coherent sources with different average photon numbers $\bar{n}_1$ and $\bar{n}_2$.  Alice determines the percentage of each of these states that is sent down the channel.  If Alice sends Bob a total of 100 pulses, of which 70 (70\%) have an average photon number of $\bar{n}_1$ and 30 (30\%) have $\bar{n}_2$ and we assume a loss of 50\%, then Bob should receive 35 (70\%) pulses with an average photon number of $\bar{n}_1$ and 15 (30\%) with $\bar{n}_2$.  In this scenario, we define loss as losing the whole pulse.  Loss affects the total number of photons received, but not the percentage of $\bar{n}_1$ and $\bar{n}_2$.  Eve performs a PNS attack by replacing all or part of the lossy transmission line with a lossless line and altering the percentage of $\bar{n}_1$ and $\bar{n}_2$ sent through to Bob.  In this example we assume Eve has replaced the entire transmission line with a lossless one.  Eve sits on the line and measures number until she finds a pulse containing more than one photon and then she takes one of these photons and lets the other pass.  She blocks enough of the single photon pulses such that the initial loss is preserved.  If $\bar{n}_1 < \bar{n}_2$, The $\bar{n}_2$ pulse will have more photons on average than the other and therefore will be allowed to pass through to Eve more than the other.  So, in the presence of Eve, if Alice sends 100 pulses, of which 70 (70\%) have an average photon number of $\bar{n}_1$ and 30 (30\%) have $\bar{n}_2$ and we assume a loss of 50\% which Eve will take over, then Bob would still receive a total of 50 pulses, but the percentages of $\bar{n}_1$ pulses will be less than $30\%$ and  the percentage of $\bar{n}_2$ pulses will be greater than $70\%$, which is not identical to what Alice sent.  Here, we are looking at the very worst possible case of eavesdropping.  We are assuming that Eve has replaced all of the noise with a noiseless channel.

\begin{figure}[ht]
	\centering
	\includegraphics[scale=1]{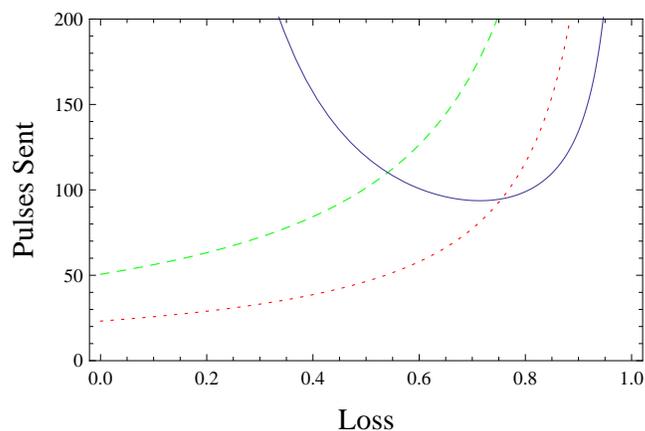}
	\caption{The solid line is the number of pulses sent by Alice (not necesarily detected by Bob) for the coherent decoy state scheme to detect an eavesdropper with a 99\% confidence.  The dotted and dashed lines are for the EE BB84 scheme at 10\% and 30\% dephasing respectively.  For cases of very high loss, decoy states outperform EE BB84.  However,  for more moderate levels of loss, EE BB84 requires fewer pulses to confidently detect the presence of an eavesdropper compared to coherent decoy states.}
	\label{fig:D_number_vs_loss}
\end{figure}

Alice looks at the percentage of $\bar{n}_1$ and $\bar{n}_2$ received by Bob and compares it to the percentages she sent.  If she can tell the difference between them with an acceptable confidence, then Eve is detected.  This is treated in the same way we treated EE BB84 above.  The Chernoff distance will give us a metric to determine the presence of Eve and the number of pulses needed to be 99\% confident of the presence of an eavesdropper is given in figure \ref{fig:D_number_vs_loss}.  The efficiency of coherent decoy states improves as loss rises because it gives Eve more space to sift the photons, but as the loss becomes too high, then obviously transmission becomes difficult for any scheme.

\section{Conclusion}

The crux of the coherent decoy state solution is that Eve manipulates photon number statistics in a way that Alice can detect.  However, if Eve can gain information, which allows her to not alter the statistics in a detectable manner, then the coherent decoy state technique will not be a successful solution.  This situation would obviously justify the implementation of EE B84, yet EE BB84 is advantageous in some other scenarios as well.

The parameters and performance of EE BB84 and coherent decoy states can vary greatly depending on environment and choice of variables.   For the examples in figure \ref{fig:D_number_vs_loss}, the coherent decoy state parameters were chosen such that the percentage of $\bar{n}_1$ pulses is $70\%$ and the percentage of $\bar{n}_2$ pulses is $30\%$, and the dephasing for the EE BB84 scheme was set to $10\%$ and $30\%$ for the two lines respectively.  It can be seen that for loss of less than $75\%$ and dephasing less than $10\%$ the EE BB84 scheme outperforms the coherent decoy state scheme by requiring fewer pulses.  At 50\% loss the EE BB84 scheme would need to send about a third the number of pulses as the coherent decoy state to detect an eavesdropper with 99\% confidence.

Coherent decoy states are a popular solution to the photon number splitting attack for a reason.  They achieve linear scaling with transmittivity.  Additionally, coherent decoy states can be used to distill a secret key without Bob alternating detection modes.  However, in EE BB84, Bob must alternate between a polarization detection mode and a phase detection. This gives coherent decoy states an advantage over the present version of EE BB84.

At the moment, EE BB84 does not possess general superiority to coherent decoy states.  Therefore, the appeal of EE BB84 is that it has some situational advantages and approaches the problem of the photon number splitting attack in a manner strategically different from that of coherent decoy states.  The general strategy of coherent decoy states is to improve the secret key transmission rate by focusing on limiting the amount of information that Eve can possibly obtain while still avoiding detection.  Meanwhile, the strategy behind EE BB84 is direct detection of an eavesdropper that might be performing quantum non-demolition measurements.  The strategy of EE BB84 is not superior to that of coherent decoy states.  It is simply different, and this difference helps generate situations where the EE BB84 scheme has specific advantages, like the case when the operation time for the key transmission is not long enough for decoy states to be a robust defense.  In cases such as this, EE BB84 has an advantage because of its ability to determine the use of quantum non-demolition measurement with a rather meager number of pulses.

\end{document}